\documentclass[twocolumn,showpacs,preprintnumbers,amsmath,amssymb,superscriptaddress]{revtex4}
\usepackage{epsfig,bm}
\begin{document}
%\preprint{APS/123-QED}
\title{Inversion doublets of $3N$+$N$ cluster structure in 
excited states of $^4$He} 
\author{W. Horiuchi}%
\affiliation{Graduate School of Science and Technology, 
Niigata University, Niigata 950-2181, Japan}
%\email{horiuchi@nt.sc.niigata-u.ac.jp}
\author{Y. Suzuki}
%\altaffiliation{Corresponding author.}
\affiliation{Department of Physics, and Graduate 
School of Science and Technology, Niigata University, Niigata
950-2181, Japan}
%\email{suzuki@nt.sc.niigata-u.ac.jp}

\begin{abstract}
Excited states of $^4$He are studied in four-body calculations with
explicitly correlated Gaussian bases. All the levels 
below $E_{\rm x}$=26\,MeV are reproduced reasonably well 
using realistic potentials. 
An analysis is made to show how 
the $0^+_2$ state becomes a resonance but those having almost 
the same structure as this state in different spin-isospin channels 
are not observed as resonances. The role of tensor force is 
stressed with a particular attention to the level 
spacing between the two $0^-$ states.
The calculation of spectroscopic 
amplitudes, nucleon decay widths, and spin-dipole transition 
strengths demonstrates that the $0^+_2$ state and the three lowest-lying  
negative-parity states with $0^-$ and $2^-$ have $^3$H+$p$ 
and $^3$He+$n$ cluster configurations, leading to the 
interpretation that these negative-parity states are 
the inversion doublet partners of the $0^+_2$ state. 
\end{abstract}
\maketitle

\section{Introduction}
The competition of particle-hole and cluster excitations is 
one of the most interesting issues in the structure of light nuclei. 
They emphasize different aspects of nuclear excitation modes, and 
often coexist in low-lying spectrum. 
Both the excitations are usually described in quite different 
languages, thus defying the reproduction of such a coexistence in a 
single scheme. 
In fact some intruder states have remained unresolved even 
in a large-space calculation based on realistic interactions.  
For example, the excitation energy of the 
so-called Hoyle state which is recognized to have 
large overlap with 3$\alpha$ configuration~\cite{neff}  
is predicted too high in the no-core shell model~\cite{no-core}. 
According to the shell model, negative-parity states should appear first 
in the excited spectrum of $^{16}$O, but they show up just above the 
first excited $0^+$ state which  
is also understandable from $^{12}$C+$\alpha$ structure~\cite{suzuki}. 

The $^{4}$He nucleus is a lightest system offering a
coexisting spectrum. Its ground state is doubly magic 
and tightly bound, but its first excited state 
is not a negative parity but $0^+$, similarly 
to $^{16}$O. This state was first conjectured as 
a breathing mode, but an extensive study has confirmed it   
as a cluster state of $3N$+$N$ ($^3$H+$p$ and $^3$He+$n$) 
configuration~\cite{hiyama}.  
Accepting this interpretation for this state, we are led to 
the following questions. Since the $3N$ and $N$ clusters 
having spin 1/2 and isospin 1/2 
move in a relative $S$ wave, four states may appear 
which all have basically the same $3N$+$N$ configuration but 
different $J^{\pi}T$ with $0^+0, 1^+0, 0^+1, 1^+1$. These states 
may be called quadruplets. The first question we set here is 
`Why do we actually observe only one of them, $0^+0$?' 

The second question is concerned with the concept of an 
inversion doublet which is known in molecular spectroscopy. 
For a system consisting of asymmetric molecules 
(clusters), one may expect a partner state of negative parity as in 
the ammonia molecule. These positive- and negative-parity pairs are 
called inversion doublets. 
On the analogy of the molecule we may ask a question 
`What about the possibility of observing 
negative-parity partners in which the $3N$ and $N$ clusters 
move in a relative $P$ wave?' The negative-parity partners 
would have $J^{\pi}$=$0^-,\,1^-$, and/or $2^-$, which result from 
the coupling of the spins of the two clusters and the relative 
orbital angular momentum between them. The 
centrifugal barrier for the $P$ wave is 
more than 3\,MeV at the $3N$-$N$ relative distance of 4\,fm,  
so that the expected partner states may   
appear in the region of the excitation energy 
$E_{\rm x}$=21-23 MeV. In fact 
the $0^-$ and $2^-$ states are observed in this region. 
Traditionally, these states are considered 
$s^3p$ shell-model states, but could be better understood from the 
$3N$+$N$ configuration. The ATMS variational calculation seems to 
suggest this picture for the negative-parity states~\cite{atms} but no discussion 
was made on the relationship between them and the $0^+_20$ state. 
According to recent large-space shell-model calculations, these $0^{\pm}0$ states  
show quite different convergence~\cite{navratil}: Because of its slow convergence, 
the $0^+_20$ state is attributed to a radial excitation. 

The purpose of this study is to 
answer the two questions by performing four-body calculations with 
realistic potentials. Thus we are mainly  
interested in the three 
excited states, $0^+0$\,($E_{\rm x}$=20.21\, MeV), $0^-0$\,(21.01\, MeV), 
and $2^-0$\,(21.84\, MeV), but also consider other excited 
states which all have a width larger than 5\,MeV. 
We will not invoke any model ansatz,  
that is, our calculation is based on neither the shell model 
nor an RGM calculation~\cite{csoto, hofmann} 
which couples $^3$H+$p$, $^3$He+$n$, and $d$+$d$ two-cluster 
channels, but treat four nucleons equally in an unconstrained 
configuration space. We will obtain the energies and wave functions 
of the excited states of $^4$He 
in a basis expansion method. The basis used here 
is square integrable, so that the excited states are obtained 
in a bound-state approximation. As we show later, this approximation 
works fairly well for predicting the three lowest-lying excited states, 
but it gives only a qualitative prediction for the other broad 
levels.  

Section~\ref{formulation} gives a brief description of  
the basis functions used to solve the four-body problem.  
Section~\ref{result} presents the results of calculation together with 
some discussions. We show the energy spectrum of $^4$He in 
Sec.~\ref{spectrum}, discuss the problem about the quadruplets in 
Sec.~\ref{quadruplets} and  answer the 
question concerning the inversion doublets in Sec.~\ref{negative-parity}.  
Section~\ref{conclusion} draws a conclusion of the present work.

\section{Formulation}
\label{formulation}
The Hamiltonian $H$ for a system of two protons and two neutrons 
consists of the 
kinetic energy ($T$) and a nucleon-nucleon 
potential including the Coulomb potential ($V_{\rm Coul}$). 
The center of mass kinetic 
energy is properly subtracted. 
A three-body force is ignored as it has a small effect on the 
spectrum above the $3N$+$N$ threshold~\cite{carlson}. 
We use the G3RS potential~\cite{tamagaki} and the AV8$^\prime$
potential~\cite{av8} 
as the two-nucleon interaction. 
Both of them contain central ($V_{\rm c}$), tensor ($V_{\rm t}$) 
and spin-orbit ($V_{\rm b}$) terms. The ${\bm{L}}^2$ and 
${\bm L}\!\cdot\!{\bm S}$ terms of the G3RS potential are ignored.  
The ground-state properties of $d$, $^3$H, $^3$He, and $^4$He given by 
these potentials are similar to each other~\cite{DGVR}. 
The tensor and spin-orbit forces of the AV8$^\prime$ potential are, however, stronger 
than those of the G3RS potential, while the central force of the AV8$^\prime$ potential 
is weaker than the one of the G3RS potential. 

A variational solution $\Psi_{JM_JTM_T}$ for the Schr\"{o}dinger equation 
is obtained by taking a linear combination of many basis states, 
each of which has the following form
\begin{align}
&\Phi_{(LS)JM_JTM_T}\notag\\
&=\mathcal{A}\big\{
\text{e}^{-\frac{1}{2}\tilde{\bm{x}}A\bm{x}}
\left[\left[\mathcal{Y}_{L_1}\!(\widetilde{u_1}\bm{x})
\mathcal{Y}_{L_2}\!(\widetilde{u_2}\bm{x})\right]_L\!\chi_S\right]_{JM_J}
\!\eta_{TM_T}\big\},\label{basis}
\end{align}
where $\mathcal{Y}_{\ell}(\bm{r})=r^{\ell}Y_{\ell}(\widehat{\bm{r}})$ is a 
solid spherical harmonic. 
Here $\mathcal{A}$ is the antisymmetrizer, 
$\bm{x}$ a column vector whose elements are 
three relative coordinates $(\bm{x}_1, \bm{x}_2,\bm{x}_3)$, and  
$A$ is a 3$\times$3 positive-definite, symmetric matrix whose 
6 independent elements are variational parameters.  
The vectors  $u_1$ and $u_2$ each 
contain three elements determining the weightings of the 
relative coordinates and are used to  
specify the angular motion of the basis (\ref{basis}). 
The tilde $\tilde{\ }$ stands for the transpose
of a column vector, and thus 
the inner product $\widetilde{u_1}\bm{x}$, which we call a global vector, 
is a vector 
in three-dimensional coordinate space. However, the inner product 
$\tilde{\bm{x}}A\bm{x}$ denotes a scalar in three-dimensional 
space as it is defined by $\sum_i\bm{x}_i \!\cdot\!(A\bm{x})_i$=
$\sum_{i,j}A_{ij}\bm{x}_i\!\cdot\!\bm{x}_j$.  

The global vector representation for the rotational 
motion used in Eq.~(\ref{basis}) is found to be very useful. 
A reader is referred to Refs.~\cite{DGVR,usukura} for more detail. 
The spin 
function $\chi_{SM_S}$ in Eq.~(\ref{basis}) 
is specified in a successive coupling,  
$[[[\tfrac{1}{2}\tfrac{1}{2}]_{S_{12}}\tfrac{1}{2}
]_{S_{123}}\tfrac{1}{2}]_{SM_S}$, and all possible intermediate 
spins $(S_{12}, S_{123})$ are taken into account in the calculation. 
The isospin function $\eta_{TM_T}$ is also treated in exactly the 
same way as the spin function. We include the following ($LS$) 
values in Eq.~(\ref{basis}) to 
obtain the state with $J^{\pi}$ for both $T$=0 and 1:  
\begin{align*}
J^{\pi}&\quad (LS) \\
0^+&\quad (00), (22)\,;\, (11)\\
1^+&\quad (01), (21), (22)\,;\, (10), (11), (12), (32)\\
0^-&\quad (11)\,;\, (22)\\
1^-&\quad (10), (11), (12), (32)\,;\, (21), (22)\\
2^-&\quad (11), (12), (31), (32)\,;\, (20), (21), (22), (42).
\end{align*}
Here the semicolon divides a natural parity set from 
an unnatural parity one. The values of $L_1, L_2$ in Eq.~(\ref{basis})  
for a given $L$ are chosen to be $L, 0$ for the natural parity 
and $L, 1$ for the unnatural parity. 
Any basis functions with $L^{\pi}$=$0^-$ are not included in the  
present calculation. 

Each basis function differs in the choices of $A,\, u_1$, and $u_2$.  
The exponential part specified by $A$ is called an explicitly 
correlated Gaussian. An alternative expression for this part is given 
using the single-particle coordinate $\bm{r}_i$ as~\cite{book} 
\begin{align}
\text{e}^{-\frac{1}{2}\tilde{\bm{x}}A\bm{x}}
=\exp\Big[-\frac{1}{2}\sum_{i<j}\Big(\frac{\bm{r}_i-\bm{r}_j}{b_{ij}}\Big)^2
\Big].
\label{ECG}
\end{align} 
Specifying the elements of $A$ using 
the 6 variables, $(b_{12},\,b_{13},\,\ldots, \,b_{34})$, 
is convenient for controlling 
the spatial extension of the system.

\section{Results}
\label{result}
\subsection{Energy spectrum}
\label{spectrum}

The accuracy of our solution depends on the basis 
dimension and the optimization of the variational parameters. 
The selection of the parameters is performed by the 
stochastic variational method~\cite{svm,book}. 
As all the states but the ground state are resonances, 
increasing the basis size unconditionally does not always lead to a 
solution we are seeking. Namely, if the 
variational parameters are allowed to reach very far in the spatial 
region, the energy for the excited state would fall down to the $^3$H+$p$ 
threshold. 

\begin{figure}[t]
\epsfig{file=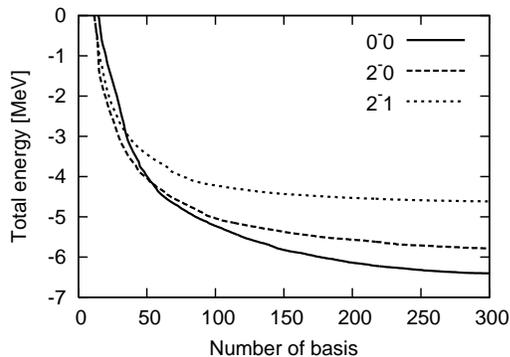,scale=1.1}
\caption{The energy convergence of the three lowest-lying negative-parity 
states of $^4$He calculated using the G3RS potential.}
\label{E-conv.fig}
\end{figure}

Some details of the calculation are given below. 
The $b_{ij}$ parameters are restricted to $0\!<\!b_{ij}\!<\!8$\,fm for 
all the states but the $0^-1$ state. This choice covers the configuration space 
large enough to obtain accurate solutions for both the ground and 
first excited states~\cite{DGVR}.
Each element of $u_i$ is allowed to take 
a value in the interval $[-1,1]$ under the constraint that its norm 
is unity, ${\it i.e.}$ $\widetilde{u_i}u_i$=1. Note that changing the 
normalization of $u_i$ does not actually alter the basis function (\ref{basis}) 
except for its normalization. We found that using 600 basis states 
of the form given in Eq.~(\ref{basis}) (that is, 600 choices of parameters 
for $A,\,u_1,\,u_2,\,L,\,S,\,S_{12},\,S_{123},\,T_{12},\,T_{123}$) 
enabled us to obtain converged solutions for both 
the ground and first excited states. See Fig. 1 of Ref.~\cite{DGVR}. 
Solutions for the other states are obtained in the basis dimension of 300. 
Figure~\ref{E-conv.fig} displays the energy convergence of the 
three lowest-lying negative-parity states with $J^{\pi}T$=$0^-0,\, 2^-0$, 
and $2^-1$ as a function of basis dimension. The energies of 
these states are considerably stable for the increase of basis 
functions, though they do not have 
proper asymptotic behavior characteristic of a resonance. The energies 
of the other levels are not very accurately obtained. They have a large 
width, so that we think our calculation gives only an approximate 
energy. In particular we found that the energy of the $0^-1$ state, which has  
a width of about 8\,MeV, was not as stable as the other states. 
We thus obtained its energy by restricting the range of 
$b_{ij}$ as $0\!<\!b_{ij}\!<\!6$\,fm.

\begin{figure}[b]
\epsfig{file=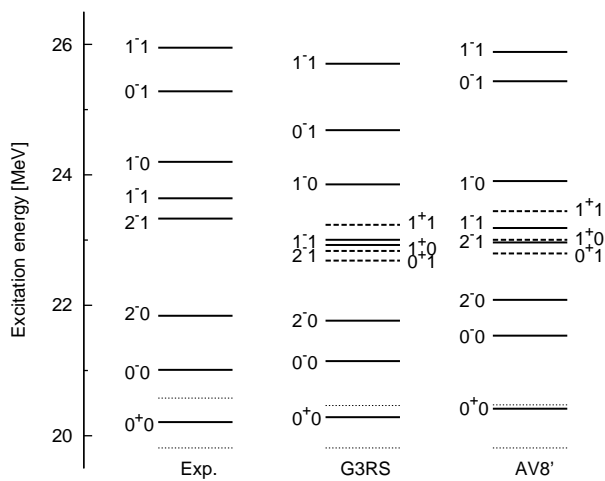,scale=1}
\caption{Energy levels of the excited states of $^{4}$He labeled with 
$J^{\pi}T$. Three of the quadruplets are drawn by dashed lines. 
The dotted lines indicate the $^3$H+$p$ and $^3$He+$n$ thresholds, 
respectively. Experimental 
values are taken from Ref.~\cite{tilley}.}
\label{spectrum.fig}
\end{figure}

Figure~\ref{spectrum.fig} compares the 
spectrum of $^{4}$He between theory and experiment.  
The calculated binding energy of $^3$H is 7.73\,MeV for G3RS and 
7.76\,MeV for AV8$^\prime$~\cite{DGVR}. Thus the calculated 
$^3$H+$p$ threshold energy does not agree by about 0.7\,MeV 
with the experimental one. 
The theoretical spectrum in the figure is drawn by shifting the calculated 
energies downward so as to adjust this difference of the threshold energy.  
The theory reproduces the level sequence of the spectrum as a whole 
and especially the excitation energies 
of the $0_2^+0$, $0^-0$, and $2^-0$ states very well. 
The levels above $E_{\rm x}$=23\,MeV are predicted 
to be slightly lower than 
experiment except for the $0^-1$ level with AV8$^\prime$. 
As their widths are all larger than 5\,MeV, this discrepancy may  
be allowable in the bound-state approximation for unbound states.  
Noteworthy is that the 
calculation predicts three states with $0^+1$, $1^+0$, and $1^+1$ 
around $E_{{\rm x}}$=23\,MeV, as denoted by dashed lines. 
These states together with the $0^+_20$ state are the quadruplets relevant to 
the first question. As speculated, they show up in the present  
calculation, but no such states are observed experimentally.

\subsection{Quadruplets}
\label{quadruplets}

\begin{table}[b]
\caption{Percentages of $(LS)$ components of the quadruplets calculated  
using the G3RS potential. Unnatural 
parity components are negligibly small.}
\begin{tabular}{cccccccccccccc}
\hline\hline 
     && $\ 0_2^+0\ $ & $\ 0^+1\ $ && $\ 1^+0\ $ & $\ 1^+1\ $\\
\hline
(00) && 93.0 & 93.4  && --   & --\\
(01) && --   & --    && 93.3 & 93.4 \\
(21) && --   & --    && 3.0  & 3.4 \\
(22) && 6.9  & 6.6   && 3.7  &3.1 \\
\hline\hline
\end{tabular}
\label{decom+.table}
\end{table}

To resolve the first problem on the quadruplets, we begin by understanding why 
only the $0^+_20$ state gets considerably lower than the 
other quadruplet members. 
As shown in Table~\ref{decom+.table}, all the members of 
the quadruplets consist of about 93\% 
$L$=0 and 7\% $L$=2 components. These values are 
almost equal to the corresponding components in $^3$H and $^3$He~\cite{DGVR}, 
consistently with  
the conjecture that the quadruplets have $3N$+$N$ cluster structure with 
a relative $S$-wave motion.

\begin{table*}[t]
\begin{center}
\caption{Total energy of the quadruplets, given in MeV, and
its decomposition into the contributions from the kinetic energy 
and the different potential pieces. The row and column 
of each 3$\times$3 matrix correspond to the configuration space with 
$L$=0,\,2, and 1. 
The off-diagonal elements in the lower triangular part of the matrix 
are added to the corresponding elements in the upper triangular part. For example,
4.58, $-$22.65 and $-$0.00 in the first row means 
the energy contribution of $(L, L^\prime)$=(0,0) channel,
(0,2) and (2,0) channels and (0,1) and (1,0) channels, respectively.
The contributions of $L$=3 
are negligible and omitted. The G3RS potential is used.}  
\label{decompot+.table}
\begin{tabular}{lcccccccccccccccc}
\hline\hline
&\multicolumn{7}{c}{$T$=0}&&\multicolumn{7}{c}{$T$=1}\\
\cline{2-8}\cline{10-16}
&\multicolumn{3}{c}{$0^+$}&&\multicolumn{3}{c}{$1^+$}&
&\multicolumn{3}{c}{$0^+$}&&\multicolumn{3}{c}{$1^+$}\\
\cline{2-4}\cline{6-8}\cline{10-12}\cline{14-16}
 &4.58&$-$22.65&$-$0.00&&6.48&$-$21.74&$-$0.01&&6.30&$-$21.67&$-$0.00&
&6.62&$-$21.31&$-$0.01\\
$\left<H\right>$&    &10.97&$-$0.29   &&    &10.64&$-$0.16   &&    &10.58&$-$0.12 &
&    &10.45&$-$0.15   \\
   &    &     &0.14      &&    &     &0.08      &&    &     &0.06 &
&    &     &0.08      \\
\cline{2-4}\cline{6-8}\cline{10-12}\cline{14-16}
                &29.26&--&--     &&31.09&--&--     &&31.45&--&--     &
&31.73&--&--     \\
$\left<T\right>$&     &10.99&--  &&     &10.30&--  &&     &10.31&--  &
&     &10.08&--  \\
                &     &     &0.15&&     &     &0.08&&     &     &0.06&
&     &     &0.08\\
\cline{2-4}\cline{6-8}\cline{10-12}\cline{14-16}
 &$-$25.07&--&--&&$-$25.00&--&--&&$-$25.54&--&--&
&$-$25.50&--&--\\
$\left<V_{\rm c}\right>$&  &$-$1.56&-- &&  &$-$1.26&-- &&  &$-$1.28&--&
&  &$-$1.17&-- \\
                        &  &  &$-$0.01 &&  &  &$-$0.00 &&  &  &$-$0.00&
&  &  &$-$0.00 \\
\cline{2-4}\cline{6-8}\cline{10-12}\cline{14-16}
                            &0.39&--&--&&0.39&--&--&&0.39&--&--&&0.40&--&--\\
$\left<V_\text{Coul}\right>$&  &0.03&--&&  &0.03&--&&  &0.03&--&&  &0.02&--\\
                            &  &  &0.00&&  &  &0.00&&  &  &0.00&&  &  &0.00\\
\cline{2-4}\cline{6-8}\cline{10-12}\cline{14-16}
                        &--&$-$22.65&-- &&--&$-$21.75&--&&--&$-$21.67&-- &
&--&$-$21.31&-- \\
$\left<V_{\rm t}\right>$&  &1.54&$-$0.30&&  &1.61&$-$0.16&& &1.56&$-$0.13&
&  &1.55&$-$0.13\\
                        &  &    &0.01   &&  &    &0.00   &&  & &0.00   &
&  &    &0.00   \\
\cline{2-4}\cline{6-8}\cline{10-12}\cline{14-16}
                        &--&--&$-$0.00  &&--&--&$-$0.01 &&--&--&$-$0.00  &
&--&--&$-$0.01  \\
$\left<V_{\rm b}\right>$&  &$-$0.02&0.00&&  &$-$0.03&0.00&& &$-$0.05&0.00&
&  &$-$0.03&0.00\\
                        &  &       &0.00&&  &       &0.00&&  & &0.00&
&  &       &0.00\\
\hline\hline
\end{tabular}
\end{center}
\end{table*}

We list in Table~\ref{decompot+.table} the energy contents of the 
quadruplet members.  The matrix elements of the Hamiltonian $H$ as well  
as its every piece are shown in a matrix form. The row and column labels  
of the 3$\times$3 matrix correspond to the configuration space 
with $L$=0, 2, and 1. The off-diagonal 
elements in the lower triangular part of the matrix are added to the 
corresponding elements in the upper triangular part.   
We see that the key elements which give only the 
$0^+_20$ state about 3\,MeV larger binding energy than the other members 
are the kinetic energy as well as the tensor force. The kinetic 
energy contribution from the main channel listed in Table~\ref{decom+.table} 
is found to be about 2\,MeV smaller in the $0^+_20$ state 
than in the other states. This is a consequence of   
the symmetry of the orbital part of the wave function as understood from 
Wigner's supermultiplet theory~\cite{wigner}.  
The spin and isospin function of four nucleons contains more number of 
antisymmetric pairs in $S$=0, $T$=0 channel than in other $ST$ channels, so 
that the orbital part of the $0^+_20$ state is more 
symmetric with respect to the nucleon-exchange than the other 
orbital functions. Furthermore, 
the $0^+_20$ state gains about 1\,MeV energy 
compared to the others states, due to the tensor coupling between the 
$L$=0 and $L$=2 components. 

\begin{figure}[b]
\epsfig{file=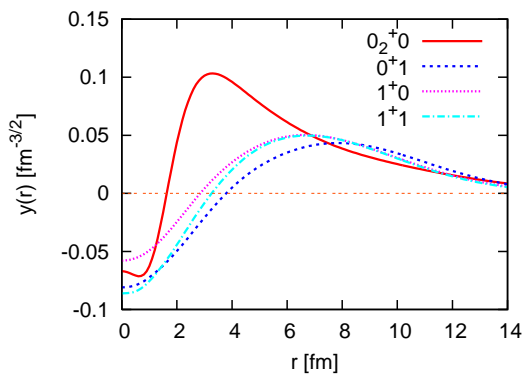,width=7.2cm,height=5.04cm}
\caption{(Color online) SAs of the quadruplets 
for the $S$-wave $^3$He+$n$ decay. The G3RS 
potential is used. }
\label{RWA_plus.fig}
\end{figure}

Now we discuss whether or not the quadruplets can be 
observed as resonances in $^3$H+$p$ and 
$^3$He+$n$ decay channels. To this end we calculate 
a spectroscopic (or reduced width) amplitude (SA) defined as  
\begin{align}
y(r)&=
\sqrt{\tfrac{4!}{3!}}\Big<
\left[\left[\Psi_{\frac{1}{2},\frac{1}{2}m_t}(3N)
\phi_{\frac{1}{2},\frac{1}{2}\,-m_t}(N)
\right]_{I}Y_{\ell}(\hat{\bm{R}})\right]_{JM_J}
\notag\\
&\times\frac{\delta(R-r)}{Rr}
\Big|\Psi_{JM_JT0}(^{4}\text{He})\Big>.
\label{RWA.eq}
\end{align}
Here $\bm{R}$ is the $3N$-$N$ relative 
distance vector, $\Psi_{{1}/{2},{1}/{2}\,m_t}$ 
the normalized 3$N$ ground-state wave function, and 
$\phi_{{1}/{2},{1}/{2}\,-m_t}$ is the 
nucleon spin-isospin function. They are coupled to the channel spin $I$. 
The label $m_t$ distinguishes either $^3$H+$p$ 
($m_t$=$1/2$) or $^3$He+$n$ ($m_t$=$-1/2$) channel. 
The 3$N$ wave function $\Psi_{{1}/{2},{1}/{2}\,m_t}$ used here 
is obtained in the calculation using 
the basis of type (\ref{basis}) 
with $(LS)$=$(0\frac{1}{2})$ and $(2\frac{3}{2})$~\cite{DGVR}. 
The component of $L$=1 is very small (0.05\%) and ignored unless otherwise. 
Figure~\ref{RWA_plus.fig} displays the $^3$He+$n$ SA 
of the quadruplets. The orbital angular momentum between 
the clusters 
is set to $\ell$=0, and so $I$ is equal to $J$. 
The $^3$H+$p$ SA is virtually the same as the $^3$He+$n$ SA 
(except for the phase). 
The $0^+_20$ state exhibits behavior quite different 
from the others: Its peak position is 3\,fm, outside 
the $3N$ radius ($\sim$2.3\,fm). Moreover, the spectroscopic factor   
defined as $\int_0^\infty y^2(r)r^2 dr$ is so large as 1.03. 
In a sharp contrast to the $0^+_20$ state, 
the SAs of the 
other quadruplet members show nothing of resonant behavior: 
The peaks are located extremely far outside the $3N$ radius, and the 
$y^2$ value is small in the inner region. In passing we note that 
the SA of the ground state has a sharp contrast with 
that of the $0^+_2$ state:  The peak appears near 
the origin and the amplitude is confined mostly in the 3$N$ radius. 

Since our variational solution is expected to be fairly accurate at least 
in the inner region, a decay width can be estimated with the formula 
of $R$-matrix type: 
\begin{align}
\Gamma_N=2P_{\ell}(kr)\frac{\hbar^2 r}{2\mu}y^2(r),
\label{width}
\end{align}
where $k$ is the wave number given by $k$=$\sqrt{2\mu E/\hbar^2}$ with 
the decay energy $E$, $\mu$ the reduced mass of the decaying particles, 
and $P_{\ell}$ is the penetrability 
\begin{align}
P_{\ell}(kr)=
\begin{cases}
\frac{kr}{F_{\ell}^2(kr)+G_{\ell}^2(kr)} \quad 
&\text{for}\quad ^3\text{H}\!+\!p\\
\frac{kr}{(kr)^2[j_{\ell}^2(kr)+n_{\ell}^2(kr)]}\quad
&\text{for}\quad ^3\text{He}\!+\!n
\end{cases}
\end{align}
which is expressed in terms of either Coulomb wave functions or 
spherical Bessel functions.  
The decay width (\ref{width}) depends on the 
channel radius $r$, but its dependence is found to be 
mild: The $\Gamma_p$ value 
of the $0^+_20$ state is 0.69, 0.74, 0.67\,MeV at $r$=4, 5, 6\,fm, 
in good agreement with the empirical value 
of 0.50\,MeV~\cite{tilley}.  
The above analyses all confirm that the $0^+_20$ state has well-developed 
$3N$+$N$ cluster structure, in accordance with the conclusion of 
Ref.~\cite{hiyama}. Moreover, we conclude 
that none of the quadruplets except for the $0^+_20$ state is a physically 
observable resonance. This conclusion is consistent with the RGM phase-shift 
analysis which finds no resonance around 23\,MeV 
excitation energy region\cite{hofmann}.

As discussed above, all the quadruplet members but the $0_2^+0$ state 
do not gain energy large enough to come down below the $^3$He+$n$ threshold. 
Only the $0_2^+0$ state shows up between the $^3$H+$p$ 
and $^3$He+$n$ thresholds thanks to their Coulomb energy 
difference. The isospin conservation gives the 
$0_2^+0$ state an almost equal mixing of the open ($^3$H+$p$) 
and closed ($^3$He+$n$) channels. 
Both effects of the 
isospin conservation and the $^3$H+$p$ Coulomb barrier make 
the $\Gamma_p$ value of the $0_2^+0$ state rather small. This state 
is thus a good example of a Feshbach resonance~\cite{feshbach}.

\subsection{Negative-parity partners of the first excited 0$^+$ state}
\label{negative-parity}

Before coming to the inversion doublet issue, we first comment on the features of 
the negative-parity states. 
According to the shell model, the negative-parity states basically arise from 
the $s_{1/2}^{-1}p_{3/2}$ or $s_{1/2}^{-1}p_{1/2}$ particle-hole excitation, which 
predicts $J^{\pi}$=0$^-$, 1$^-$, 1$^-$, and $2^-$ for both $T$=0 and 1. 
However, a suitable combination of the two 1$^-$ states with $T$=0 
corresponds to the excitation of the center of mass, leaving only one 
$1^-$ state with $T$=0. 
Seven negative-parity states observed experimentally below $E_{\rm x}$=26\,MeV 
include three states with $T$=0 and four states with $T$=1, 
which is in agreement 
with the shell-model prediction. However, this agreement 
may not necessarily mean that 
the negative-parity states have shell-model like structure because 
the present four-body calculation also produces seven negative-parity states, 
as shown in Fig.~\ref{spectrum.fig}. 

\begin{table}[t]
\caption{Percentage of  ($LS$) components of the negative-parity states 
calculated using the G3RS potential. The natural and unnatural parity channels
are separated by the line.}
\begin{tabular}{cccccccccccccccc}
\hline\hline
$(LS)$&&\multicolumn{3}{c}{$T$=0}&&\multicolumn{4}{c}{$T$=1}\\
\cline{3-5}\cline{7-10}
&&$0^-$&$1^-$&$2^-$&&$0^-$&$1_1^-$&$1_2^-$&$2^-$\\
\hline
(10)  &&  --&19.7& -- &&--  &51.0&42.9&--\\
(11)  &&95.5&74.2&93.0&&96.9&43.0&53.1&93.7\\
(12)  &&  --&0.8 &0.3 &&--  &0.0 &0.6 &0.2\\
(31)  &&  --& -- &2.9 &&--  &--  &--  &2.8\\
(32)  &&  --&3.4 &2.0 &&--  &4.3 &0.1 &1.7\\
\hline
(20)  &&  --& -- &0.0 &&--  &--  &--  &0.0\\
(21)  &&  --&1.8 &0.5 &&--  &1.1 &1.5 &0.5\\
(22)  &&4.5 &0.2 &1.4 &&3.1 &0.5 &1.8 &1.1\\ 
(42)  &&  --& -- &0.0 &&--  &--  &--  &0.0\\
\hline\hline
\end{tabular}
\label{decom-.table}
\end{table}

The level sequence is 0$^-$,\, 2$^-$, and $1^-$ 
in the order of increasing energy for $T$=0, while it 
is $2^-$,\,$1^-$,\,$0^-$, and $1^-$ for 
$T$=1. Therefore the energy difference between the $0^-0$ and $0^-1$ states 
becomes much larger than the one between the $1^-0$ and $1^-1$ states 
or between the $2^-0$ and $2^-1$ states. It is interesting to clarify the mechanism 
of how this large energy difference is produced compared to the 
other negative-parity states with the same $J^{\pi}$. Table~\ref{decom-.table} 
lists the percentage analysis of the seven negative-parity states 
according to their ($LS$) channels. 
The main component has $L$=1 as expected from the shell model.  
The percentages are rather similar between the states with the same 
$J^{\pi}$ but different $T$ values.  The similarity of the percentages 
is, however, not very clear  in the $1^-$ states because the values  
are fragmented into the two $1^-1$ states.  The main channel with $L$=1  
itself has a contribution from the tensor force but also gets a contribution from 
the other channels through the tensor coupling. For example, the tensor 
force couples the natural parity  channel (11) with the unnatural parity 
channel (22).  

\begin{table*}[t]
\begin{center}
\caption{Energy contents, given in MeV, 
of the negative-parity states. The G3RS potential is used.}
\label{decom0-.table}
\begin{tabular}{ccccccccc}
\hline\hline
&&$\,0^-0$&$\,0^-1$&$\,2^-0$&$\,2^-1$&$\,1^-0$&$\,1^-_11$&$\,1^-_21$\\
\hline
$\left<H\right>$ &&$-6.40$&$-2.86$&$-5.78$&$-4.62$&$-3.69$&$-4.54$&$-1.84$\\
$\left<T\right>$ &&48.38&39.10&41.08&40.25&37.72&39.30&32.48\\
$\left<V_{\rm c}\right>$&& $-28.92$&$-24.79$&$-25.71$&$-25.82$&$-23.50$&
$-25.14$&$-22.01$\\
$\left<V_\text{Coul}\right>$&&0.48&0.44&0.42&0.43&0.40&0.43&0.42\\
$\left<V_{\rm t}\right>$&&$-26.63$ &$-17.75$ &$-21.39$ &$-19.30$ &$-18.32$ &
$-19.13$ &$-12.67$ \\
$\left<V_{\rm b}\right>$&&0.29&0.14&$-0.18$&$-0.18$&0.006&0.005&$-0.06$ \\
\hline\hline
\end{tabular}
\end{center}
\end{table*}

Table~\ref{decom0-.table} lists the energy content contributed from 
each piece of the Hamiltonian. 
Most striking is a different contribution of the tensor force. 
Compared to the $0^-1$ state, 
the $0^-0$ state gains about 9\,MeV energy from the tensor force, 
while the contribution of the central force to the 
energy gain is only about its half. The energy contents given 
by the AV8$^\prime$ potential 
are similar to those of the G3RS potential: The gain by the tensor 
force is even 
larger, about 12\,MeV and the central force gives 3\,MeV difference.   
The tensor force is most attractive in the triplet even $NN$ state, and it 
can be taken advantage of  by having more number of antisymmetric 
$NN$ pairs in the isospin space. The number of such antisymmetric pairs 
is counted from 
$\langle \eta_{TM_T}|\sum_{i<j}(1\!-\!{\bm \tau}_i\!\cdot\!{\bm \tau}_j)/4|
\eta_{TM_T}\rangle$, which is $[A(A\!+\!2)\!-\!4T(T\!+\!1)]/8$ 
for $A$-nucleon system. Thus the $0^-0$ state gains more 
attraction than the $0^-1$ state through both the (11)-(11) 
diagonal and (11)-(22) off-diagonal contributions~\cite{DGVR}. 
If the unnatural parity basis were not included in the calculation, 
the $\left<V_{\rm t}\right>$ value of the $0^-0$ state would decrease to about 
half~\cite{DGVR} and the $0^-0$ state would lose significant energy.
The role of the tensor force in lowering the energy of the $0^-0$ state 
was discussed many years ago~\cite{atms,barrett}. 
To be exact, the energy difference between the two 
states is actually a combined effect of the tensor, kinetic   
and central terms. 

\begin{figure}[b]
\epsfig{file=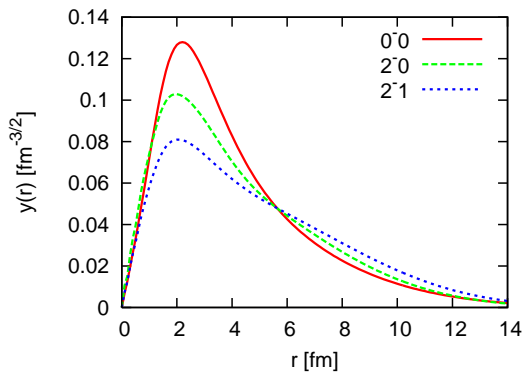,width=7.2cm,height=5.04cm}
\caption{(Color online) SAs of the three lowest-lying negative-parity states 
for the $P$-wave $^3$He+$n$ decay with $I$=1. The G3RS 
potential is used.}
\label{RWA_minus.fig}
\end{figure}

Now we discuss the characteristics of the low-lying negative-parity states 
from the viewpoint of clustering.  
In Fig.~\ref{RWA_minus.fig} we display the $^3$He+$n$ SAs calculated from the 
three lowest-lying negative-parity states with $0^-0$, $2^-0$, and $2^-1$. The 
$\ell$ value for the $^3$He-$n$ relative motion is 1, and the channel 
spin $I$ is 1.   
As expected, each of the three curves shows behavior suggesting  
$3N$+$N$ cluster structure: The peaks are centered 
around 2\,fm near the $3N$ surface, and the 
$y^2$ values are fairly large in the inner region. It is the centrifugal 
potential that makes the peak positions 
closer to the origin than that of the first excited $0^+0$ state. 
The $3N$+$N$ 
spectroscopic factors are considerably large: 0.58, 0.52, and 0.53 for the 
$0^-0$, $2^-0$, and $2^-1$ states. 
We estimate the nucleon width. Its 
channel-radius dependence is again mild, so we choose $r$=5\,fm.  
The results for ($\Gamma$, $\Gamma_p/\Gamma$), 
where $\Gamma$ is the total width in MeV, are 
(0.61, 0.72), (1.14, 0.58), and (1.85, 0.53) for $0^-0$, $2^-0$, and $2^-1$, 
respectively. These values are compared to those extracted from the 
$R$-matrix analysis~\cite{tilley}, 
(0.84, 0.76), (2.01, 0.63), and (5.01, 0.53). 
The theory predicts the width of the $0^-0$ state very well, and 
gives about half of the width for the other states.  
Though the degree of clustering is somewhat reduced in the negative-parity  
states compared to the $0^+_2$ case, 
the analysis of SA and decay width supports our conjecture that 
the $0^-0$ and $2^-0$ states as well as the $2^-1$ state constitute 
inversion-doublet partners of the first excited $0^+$ state.   
The RGM phase-shift analysis of $^3$H+$p$ scatterings~\cite{hofmann}  
supports the $P$-wave resonance interpretation 
for these negative-parity states. The SAs of the $1^-$ states around $E_{\rm x}$=24\,MeV 
show some degree of $3N$+$N$ cluster structure, though their amplitudes are considerably small 
compared to the $0^-0$ and $2^-0$ states in particular. 

An inversion doublet picture in nuclei was first proposed to understand 
the low-lying positive- and negative-parity rotation bands in $^{16}$O and 
$^{20}$Ne from an $\alpha$-core molecular structure~\cite{molecule}. 
The appearance of positive- and negative-parity partners is a natural 
consequence of the underlying intrinsic structure dominated by the existence of 
asymmetric clusters. We have shown that the three lowest-lying negative-parity 
states have a significant component of 3$N$ and $N$ clusters whose 
relative motion is in $P$-wave. It is important to realize that this result is 
obtained in the calculation which assumes no cluster ansatz for the 
wave functions. A physical reason for the appearance of the inversion 
doublet partners is that they are located near the $3N$+$N$ threshold. 

Very unique in the inversion doublets in $^4$He is that the $3N$ and 
$N$ clusters have both $J$=$1/2$, and the channel spin $I$ is different in 
the doublets: It is 0 for $0^+_20$ and 1 for $0^-0$, $2^-0$, and 
$2^-1$. The negative-parity partners with $T$=0   
should thus be characterized by the 
transition of an isoscalar spin-dipole operator, 
${\cal O}_{\lambda \nu}$=$\sum_{i=1}^4
[{\bm \sigma_i}\!\times\!({\bm r}_i\!-\!{\bm x}_4)]_{\lambda \nu}$, 
where ${\bm x}_4$ is the center of mass of $^4$He. 
Note that ${\bm r}_i\!-\!{\bm x}_4$ is proportional to 
the distance vector between nucleon $i$ and the center of mass 
of the other three nucleons. The transition 
strength to the $0^+_20$ state, $|\langle 0^+_20 ||{\cal O}_0||0^-0\rangle |^2$, 
is 11.9\,fm$^2$, which is 6.9 times larger than that to the ground state.  
Moreover, the strength  
$|\langle 0^+_20 ||{\cal O}_0||0^-0\rangle |^2$ between the doublet partners 
occupies 58\% of the ``sum rule''
$\sum_{n}|\langle 0^+_n0 ||{\cal O}_0||0^-0\rangle |^2$, where $n$ takes 
all 600 eigenstates with $0^+0$. A similar enhancement 
occurs for the $2^-0$ state as well. The value 
$|\langle 0^+_20 ||{\cal O}_2||2^-0\rangle |^2/5$ 
is 21.7\,fm$^2$, which is about 24 times larger than the one 
to the ground state, and it corresponds to 78\% of the total sum   
$\sum_{n}|\langle 0^+_n0 ||{\cal O}_2||2^-0\rangle |^2/5$. 

For the transition between the $2^-1$ and $0^+_20$ states, 
an isovector spin-dipole operator, 
${\cal O}_{\lambda \nu, 10}$=$\sum_{i=1}^4
[{\bm \sigma_i}\!\times\!({\bm r}_i\!-\!{\bm x}_4)]_{\lambda \nu}
{\tau}_{3_i}$, must be considered. The transition strength 
$|\langle 0^+_20 |||{\cal O}_{2,1}|||2^-1\rangle |^2/15$ is 17.4\,fm$^2$, 
which is 16 times larger than that to the ground state, where  
the triple bar $|||$ indicates that the reduced matrix element is taken 
in both the angular momentum and isospin spaces. 
This strength between the $2^-1$ and $0^+_20$ states 
occupies 87\% of the total strength   
$\sum_{n}|\langle 0^+_n0 |||{\cal O}_{2,1}|||2^-1\rangle |^2/15$.
   
The high collectivity of the spin-dipole strength strongly indicates 
that the intrinsic structure of 
the negative-parity states, $0^-0,\,2^-0$, and $2^-1$, is similar to 
that of the first excited $0^+_20$ state.  
 
\section{Conclusion}
\label{conclusion}
A rich spectrum of $^4$He comprising the coexisting levels 
has been reproduced in a single scheme without recourse to 
a specific model assumption. This has offered a good example of 
demonstrating the power of the global vector representation for the angular 
part of few-body systems. We have explained how only 
the $0_2^+$ state 
is observed as a resonance among the quadruplets by examining 
the symmetry property of the wave functions as well as the role of 
the tensor force. Analyzing the spectroscopic amplitudes, nucleon decay 
widths and spin-dipole transition probabilities, we 
have confirmed that both the $0_2^+$ and negative-parity states with 
$0^-0, 2^-0$, and $2^-1$ are dominated by the 
$3N$+$N$ cluster structure and that these negative-parity states 
can be understood as the inversion doublet partners of the $0_2^+$ state 
in a unified way. We have shown that the tensor force plays a vital role 
to produce the level spacing between the $0^-$ 
states with $T$=0 and 1 through the coupling between the main channel with 
$L$=1 and the unnatural-parity channel with $L$=2.  
A study of $^{16}$O in the scheme of $^{12}$C+four nucleons   
will be interesting as its spectrum has some similarity to that of $^4$He.

This work was in part supported by a Grant for Promotion of Niigata University 
Research Projects (2005-2007), and a Grant-in Aid for Scientific
Research for Young Scientists (No.\,19$\cdot$3978). W.H. is a JSPS Research 
Fellow for Young Scientists. Y.S. thanks the JSPS core-to-core 
program (Exotic Femto Systems) and the Institute for Nuclear 
Theory at the University of Washington for the support which enabled him to 
have useful communications in the INT Workshop on Correlations in Nuclei, November, 2007.

\end{document}